# GRAVITATIONAL WAVE DETECTION

## BY HOLLOW-CORE FIBER-OPTICS MACH-ZEHNDER INTERFEROMETRY

Francesco De Martini

Accademia Nazionale dei Lincei, Roma 00165. Italy

Fondazione Università Sapienza, Roma 00185, Italy

____________________________

**Abstract**: Recent advances in the field of very long distance optical communication suggest the adoption of the advanced technology based on Hollow Core Nested Anti-resonant Nodeless Fiber (HC-NANF) within the endeavour of Gravitational Wave detection using a Mach-Zehnder optical interferometer (MZ-IF). The proposal, consisting of a summary project of the device emphasizes the favorable properties of (MZ-IF) in comparison with Michelson Interferometer (M-IF) currently in operation. The key feature of the proposed method consists of the use of a couple of "fibrated" metallic antennas enfolded by a very large number ($h \times 7,7 \times 10^4$ with h= 1,2,3 etc.) of coiled rigs or of straight sections of the (HC-NANF) fiber. This amounts to a corresponding fiber length: $L^{(K)} = h \times 1500$ Km. The relevant properties of the device are noise reduction, absence of critical optical mirror alignment in a noisy environment, reduced spatial extension of the apparatus, exploration of the entire sky scenario by freely orientable antennas, a substantial cost reduction of the apparatus. The remarkable properties of (HC-NANF), invented by F. Poletti in 2013 are currently investigated by his group at the University of Southampton, UK.

## 1 - Introduction

In the last decade the successful detection of Gravitational Waves (GW) has become an established ambitious field of fundamental cosmological investigation (1). The three LIGO and VIRGO (GW) Interferometry (IF) stations based on the Michelson configuration (GW-MIF) implemented by high Q internal Fabry-Perot (FP) cavities were able to observe transient gravitational waves with a peak GW strain of $10^{-17}$ (2,3). The bulky structure of the, nearly equal, GW-MIF stations each featuring couples of liquid-He cooled, orthogonal arms extending in the land over L = 4 Km (3 Km for VIRGO) and the indispensable awkward machinery to deal with the many sources of noise together with the required extremely precise multiple IF optical alignments, suggests the adoption of a conceivably simpler optical configuration based on the modern fiber–optics (OF) technology within a different interferometry scheme. The present

francesco.demartini2@gmail.com        francesco.demartini@fondazione.uniroma1.it

work is an investigation of the pros and cons of the new proposed structure in comparison with the presently operating GM-MIF's. As we shall see, an important advance of this proposal may consist of a far limited space extension of the GW Observatory, and of no need for heavy constructions or soil excavations.

In the proposed (MZ-IF) device, the two equal OF Antennas as a very important feature of large scientific relevance, may easily act upon by two independent external motorized mechanical supports that can freely orient them in any direction towards the skies. The overall device may have a house structure similar to the one of any existing conventional Radio or Astronomical Observatory. Very likely, and perhaps somewhat relevant, the overall building cost of this apparatus is expected to be far less than that needed for the construction of a GW Interferometer similar to the ones currently in operation on the Earth.

In Chapter 2 we provide a brief description of the present GW-MIF apparatuses by analyzing some of the problems encountered in overcoming the noise and the critical alignments of the optical mirrors in a very noisy environment. In Chapter 3 we shall present our proposal based on OF with various "Variants" and "Subvariants" intended to explore a large set of the possible optical configurations of the apparatus.

2. **Noise and Locking in Michelson – Fabry - Perot Interferometers**

As it is well known, the Michelson Interferometer (MIF) consists of two single-mode monochromatic light beams travelling on two orthogonal optical paths A and B (2). First, as shown in Fig.1, in view of the typical quadrupole origin of GW radiation (3), the strict e.m. beam orthogonality that is the standard MIF configuration, does not correspond in general to all possible configurations for a complete GW detection. The two orthogonal beams, injected into one of the input ports of a single four ports 50/50 beam-splitter (BS) are linearly superimposed. The two orthogonal arms of the MIF are then terminated by two equal back high reflectivity (R=99,99%) mirrors $M_A$ and $M_B$ which reflect back on the BS the two light beams. Then there they interfere linearly giving rise at the output ports of the BS to a signal proportional to the algebraic sum of the related electromagnetic (e.m.) fields. $f_A$ and $f_B$. Specifically, if $L_A$ and $L_B$ are the spatial lengths of the beams, that is the spatial distances between $M_A$ and $M_B$ and the common BS, any length imbalance $\delta L = (|L_A - L_B|)$ is recorded at one of the output ports of the BS as a light signal owing to the corresponding phase-mismatch of the fields. Most important, In existing GW interferometers two additional equal mirrors $M'_A$ and $M'_B$, with reflectivity R ~ 85% are inserted in the two optical paths. The couples of the carefully aligned mirrors ($M_A$ - $M'_A$), ($M_B$ - $M'_B$) consist of two Fabry-Perot (FP) optical interferometers with high "finesse" F, a pure number (2). In the current operation (SW-MIF) F was in the range (500 - 1000). In virtue of the effect of the (FP) cavities the "effective length" (L) of the arms of the (MIF) are then largely increased: $L_{eff} \sim LF$. This is required to detect the very small GW spatial length perturbation: $\delta L = h L_{eff}$ where h represent the (tensorial) gravitational perturbation to the metric tensor (3) : $g_{\mu\nu} = \eta_{\mu\nu} + h_{\mu\nu}(z,t)$. There: $\eta_{\mu\nu}$ is the flat Minkowsky metric and the perturbation is be expressed as: $h_{\mu\nu}(z,t) = (h_{+\mu\nu}\, \varepsilon_{\mu\nu} + h_{X}\, \varepsilon_{\mu\nu})$ where: $h_+$ and $h_x$ are the two polarizations acting on a circular mass distribution by a propagating dilation and compression GW process along cartesian axes ($h_+$) or 45° from them: ($h_x$) [Fig.1]. The polarizations per meter (h in the range. $10^{-17} – 10^{-18}$ m: Cfr. (3)) depends on the nature and distance of the cosmic source (1,3). Within the (GW-MIF) this dilation/compression process is analyzed using a unique GW-invariant "meter": the wavelength (wl) λ of the free-space propagating e.m. field and then of the GW-MIF excitation laser radiation. The laser used in the VIRGO and LIGO (GW-IF) is a high-power solid-state single-mode Nd:YAG laser operating at a λ = 1,06 μm, stabilized by an additional thermally - insensitive reference resonance cavity (L = 40 cm). The QED technique of "optical squeezing" of the laser emitted field has been adopted to transfer a substantial amount of the QED phase-noise to the amplitude of the wave (2). In the present work we assume this complex and reliable stabilized laser system as a reference source for our project (albeit with a different λ and a different power).

The strenuous endeavour of a few highly inspired, brave people in the US, Italy and France that led to the realization of the GW-MIF lasted more than three decades, starting in 1983 from the first idea of cosmic noise-reduction by the late pisan scientist Adalberto Giazotto of the Istituto Nazionale Fisica Nucleare (INFN) (4). The pioneers of this enterprise in the US where: Rai Weiss , Ronald Drever, Kip Thorne and Barry Barish. In Europe: A. Giazotto, Alain Brillet, Carlo Bradaschia, Hans Kautzky and many others championed the GW endeavour for many decades (1). The real enormous problem faced by these scientists had only one name: "NOISE", sometimes also named: "SATAN", by Giazotto. (4).

We only list here some effects of the noise which have been found relevant in this context, by considering that we are concerned with spatial displacements far smaller than the atomic size.

1). Seismic noise mostly occurring at low frequencies. Sounds and infra-sounds coming from any source: earthquakes, sea waves, trains and any kind of travelling carriage or vehicle. Any kind of acoustic noise.

2). Thermal noise. This typical Planckian noise affecting the active optical and mechanical components of GW-MIF as, for instance the coatings of the mirrors. In particular, thermoelastic noise originating from the elastic strain of any metallic support or wire supporting the heavy mirrors of IF.

3). "Newtonian noise". First identified by Nicola Cabibbo, expresses the self gravitational effect due to any displacement due to GW.

**3. Hollow-Core Fiber Optics Mach-Zhender Interferometry**

According to Einstein's General Relativity any motion of matter/energy gives rise to a perturbation of the metric tensor $g_{\mu\nu}$ with is transmitted as a GW wave travelling at the velocity of light to all most remote parts of the Universe (3, 5). As said, this perturbation affects the size of any objects, of any spatial distance in the Universe with the only exception of the wavelength λ the free-space propagating light. This can then be assumed to be the invariant "meter" within the detection of any GW perturbation, as said. This is the rationale underlying the choice of Optical Interferometry for GW detection.

In the VIRGO-LIGO devices the measured length $L_{eff}$ has been selected, owing to the very large Finesse of the Fabry Perot interferometers, as large as possible in order to physically detect the overall very small effect δL. In view of the many technical problems encountered in the building and operation of the existing GW-MIF 's, the present proposal considers a simpler, conceivably cheaper, far less noise-dependent length consisting of the recently discovered, most advanced "*Hollow-Core Nested Anti-resonant Nodeless Fibre*" (HC-NANF) as a physical "very large distance" (L) that is necessarily affected by GW perturbations (6,7,8). The proposed solution has the following advantages:

a). The extended (HC-NANF) optical cable can be nested within to two localized "Antennas" that can be freely and independently oriented towards the skies, in any direction.

b). Since the ~96% of the optical power transmitted e.m. field travels mostly in air, the (HC-NANF) exhibits a very low loss for e.m. radiation at the λ =1.55 μm (C-band telecom region), an unprecedented polarization purity, and nonlinearity effects such as Stimulated Brillouin (SBS) or Raman Scattering (SRS) which are practically nonexistent.

c). These desirable qualities enable the transmission at high optical power with a very reduced noise.

We believe that the Mc-Zehnder (MZ-IF) configuration, shown in Figure 3, is conceivably far more suitable for GW detection than the more rigid Michelson (MIF) scheme. In facts, with the (MZ-IF) the use of two different field beam-splitters, ($BS_1$) and ($BS_2$) allows the propagation in the HC-NANF at a <u>reduced power</u> of the main laser field. (In the context of the present proposal we find convenient an average power in the range: 1 - 50 mW).  After the long path propagation, the two beams emerging from the (OF)'s can then be separately amplified by two ($YEr^{3+}$) high gain Optical Amplifiers before undergoing in $BS_2$ the out-of-phase linear field superposition leading to the output signal. (At the input ports of $BS_2$ we find convenient a power in the range 10 – 50 W). Using this final large field Amplification the <u>shot-noise</u> affecting the output signal can be greatly reduced.  None of these features can be implemented within the existing (GW-MIF) most rigid scheme, where a large power level (>100 KW) is <u>imposed</u> from the input main laser beam throughout the final signal detection.

We believe that these considerations provide the most important advantages of the proposed method. The present proposal has several aspects that we shall present under the form of different technical VARIANTS and sub-VARIANTS featuring different interferometric configurations and different laser sources.

**VARIANT 1.1**

The layout of simpler configuration of the proposed IF is illustrated in Fig. 3.  The pump of the overall IF is a highly stabilized, monochromatic Erbium (Er) Laser operating al a wavelength λ = 1540 nm, in the C – band where the loss of the telecom optical fibers become lowest in the entire optical communication band: Fig. 2. The intensity, frequency and phase of the laser should be carefully stabilized by an additional reference cavity similar to the system already adopted for the present (MIF) (1).

The highly monochromatic CW e.m. field is injected by a 50/50 optical beam splitter ($BS_1$) in two equal ANTENNAS A and B enfolded by a very large number $N_c$ of spires or straight sections made by the single mode (OF) with an overall large length L. The "Antenna", the real core of our proposal, shall be described in the next chapter. The field emerging at the output of the two Antennas are further amplified by a Yitterbium-Er-doped high-gain optical amplifier (EYA) and then injected into a second optical 50/50 beam-splitter ($BS_2$) were the two beams A and B are linearly superimposed.

The signal collected at one of the two output ports of $BS_2$ is proportional the <u>sum</u> of the monochromatic fields emerging from the OF's after amplification (Fig. 2). This is the "in–phase condition", where the phases of the two combined e.m. fields emerging from the Antennas are: $\psi_A = \psi_B$. The signal at the other output port of $BS_2$ expresses the <u>difference</u> between the fields. This is the "out-of-phase" condition $\psi_A = -\psi_B$. The GW detection signal is revealed at this last output port of $BS_2$. Indeed, the $BS_2$, which is identical to the single BS of any classical Michelson interferometers, including VIRGO and LIGO, is the most critical device in the apparatus. By a delicate feedback system in a "wait" operating condition (i.e. in absence of any real GW effect) the spatial position of $BS_2$ is driven by accurate piezo-electric transducers to the steady condition of <u>zero signal out</u>, i.e. a complete out-of- phase condition: $\psi_A = -\psi_B$. However the timing of the feedback time response is set such a that a sudden, transient GW excitation, lasting a fraction of a second, can be recorded by an optical pulse at the OUT port of $BS_2$. This critical device is well known and operates well in the existing GW-MIF's. Therefore we refrain from further consideration (1).

The most important and novel devices of the apparatus, representing the main point of the present proposal, are the two <u>equal Antennas</u> each of which supports a single-mode optical fiber featuring the large length L that transforms any cosmic GW perturbation into the contraction/dilation process δL leading to a measurable phase mismatch of the e.m. field δψ at each one of the two input ports of $BS_2$. In the following we shall account for the physical structure of the Antennas and then consider the structure of the Optical Fiber (OF) that we believe can better accomplish the GW detection task under consideration.

### (a) HOLLOW-CORE FIBER-OPTICS ANTENNA

Figs. 4 and 5 show the layout of the proposed Antenna. The core of the antenna may consist of a plane square slab (10 x 10 $m^2$), either bulk or hollow, made of low-weight metal (Al). A short metallic clamp emerging at the center of the square slab of the Antenna can be connected to an external tower by a motorized mechanical device allowing mutually independent orientations of the two Antennas towards the skies in any possible direction.

There are two ways to associate the selected Optical Fiber to the Antenna structure.

1) As a simpler solution, the metal Antenna slab is enfolded, in a very tight and orderly manner by a very large number $N_c$ of the coiled HCF Optical Fiber rings belonging to the (x-y) plane as represented in Fig. 4,5. In this case the transverse size of the Antenna, that is the thickness of the slab, cannot be too small in order to minimize the multiple 180° mechanical bending of the OF at the edges of the Antenna slab resulting in a severe perturbation of the field propagation in the OF and in a foreseeable heavy contribution to loss and to noise: the "bending-noise". As shown by (8) a bending radius of R >25 cm should avoid most of this problem at the expense of a large "thickness" of the Antenna slab structure (Fig. 4,5).

2) A more sophisticated (and highly recommended) solution consists by allowing the e.m. propagation within multiple, optically interconnected equal OF straight i-sections. Refer, for instance, to our Figg 4,5 representing one Antenna as a square metallic slab with surface 10 x 10 m². The active HC-OF fiber of overall length ($L = \Sigma_i L_i$) be prepared, for a single layer (h=1), in a form of a large set of equal, parallel straight i-sections of lengths: $L_i$ = 10 m disposed in an orderly, compact form, parallel to the x-axis, over one of the two plane surfaces of the square Antenna slab. At both edges of the slab, i.e. at both ends of each OF i-section, a small transparent glass 90° prism transfers the e.m. field from the circular transverse end of the (i-1)-section to the closely tangent transverse circular end of the next i-section and then to the next (i+1) etc. etc. (Fig. 6). (Each prism can be easily glued to the corresponding couple of OF ends) This procedure is repeated for all termination ends of all OF i-sections, and for the two parallel surfaces of the Antenna slab resulting in a unique e.m. field propagation path with overall length ($L^{(K)} = h \times \Sigma_i L_i$). Assuming the transverse diameter D=130 µm of the standard (HC-NANF) fiber (Fig. 6) the total length for a single layer of the active OF sections is: $L^{(1)}$ = 1500 Km. For: h=1,2,3… equal superimposed layers, the overall fiber length for each Antenna is: L = (h x 1500) Km, a length that can be made far larger than the $L_{eff}$ featured by the existing (MIF)'s, LIGO and VIRGO. This very large value of the attainable optical path length within the very localized Antenna apparatus is indeed the key feature of the proposed method.

The obvious advantages of configuration (2) are:

(a) Within each Antenna the e.m. field propagates entirely within straight sections of HC-OF. Furthermore, each 90° prism-mediated intra-sections optical coupling is virtually lossless and exactly preserves the field's phase and polarization. The overall propagation loss is then minimized and the "bending noise" is cancelled.

(b) The "thickness" and the weight of the overall supporting Antenna metal slab can be made small, a few centimeters. This can make the Antennas very manageable for mechanical support and spatial orientation.

The optical fiber most convenient for our purpose is the *Antiresonant Hollow Core Nested Nodeless Fiber* (HC-NANF) (7.8). Through resonant out-coupling high-order modes this fiber can be made to behave as a <u>single mode</u> fiber with an unprecedented polarization purity and ultralow backscattering. Furthermore, as approximately 97% of the radiation power propagates in air, nonlinearities such as Stimulated Brillouin (SBS) and Raman (SRS) scattering are negligible. In addition, the losses of the HC-NANF fiber has been recently measured to be as low as 0.22 dB/km in the Telecom C-band, less than any other type of (OF). The properties of the (HC-NANF), invented by Francesco Poletti in 2013, are currently being investigated by his group at the Electronics Research Center of the University of Southampton (UK).

The transverse structure of the HC-NANF fiber is shown in Figure 6. The transverse diameter of the bare HC-NANF is D ~ 130 μm.

**Noise:** The mechanical structure of the "fibrated" antenna suggests the presence of two possible causes of Noise. The first one is <u>acoustic – noise</u>. This can be cured by encasing each antenna within an adapted box in which air vacuum is provided. The second is <u>thermal-noise</u> optionally requiring the use of liquid - He cooling of the internal metallic slab of each Antenna.

Of course, the adoption of this new technology for GW detection comes with a series of problems that need to be solved. A dominant problem consists of the radiation absorption of the propagating CW light within the core of the very long OF. The compensation for this loss (0.22 dB/km) can be provided by a repeated insertion with intervals of ~50 km of free operation, of equal Erbium Doped Fiber Amplifiers (**EDFA**) stations (Figure 8).

### (b) ERBIUM DOPED FIBER - AMPLIFICATION (EDFA)

This is recently invented resource (9) consisting of a portion of Er doped OF which is pumped by an external semiconductor optical laser (SOL) at the wavelengths: $\lambda_p$ = 980 μm, or $\lambda_p$ =1480 μm for high-power operation. (Pumping by the SOL should not attain excessive Er-atom population inversion as to reach the laser threshold in the OF).

### c) High-power AMPLIFICATION (>500 mW)

A convenient choice for high power operation consists by a combined co-doping by Erbium and Ytterbium atoms: (Fig. 9). This co-doping is useful for eliminating the effect of mutual interactions between $Er^{3+}$ ion pairs when the doping concentration approaches the value 5.000 ppm (9, J.D. Minelli). Because such interactions are less significant for pairs of Yb ions. a high doping level (>10.000 ppm) is allowed without substantial scarification of the amplifier efficiency. Most high-power EDFA (typically 500 mW) use this configuration, and they are referred to as: EYDFA' s.

The EYDFA configuration must be implemented for the large-gain final Amplifiers placed the end of the two Antennas (Figure 3).

### VARIANT 2.0   OF - Michelson Interferometry.

**Figure 10** shows the layout of the proposed: "**ANTENNA MI-IF**". This apparatus reproduces the standard original structure of the Michelson interferometer (2) of the existing (GW-MIF).  The only variant here is that the very long Optical Fibers wrapped around the two equal antennas replace the open spaces existing between the BS and the end mirrors. This configuration requires the adoption of a 100% reflecting mirror at the output of any Antenna. An additional ~ 85% reflecting mirror placed at the input of the Antennas can reproduces the (FP) configuration of the existing GW-MIF's.

There is an interesting trick we can use.  Through an elegant optical and solid-state micro-mechanical operation, the internal "core" of the OF at the antenna <u>output</u> could be orthogonally cut and treated as to include the final 100% reflectivity as an <u>internal property</u> of the OF itself. Using the same process, acting again on the OF "core" at the antenna <u>input</u>, the Fabry-Perot ~ 85% reflectivity as an internal property of OF can be realized. This procedure greatly reduces the noise of our (OF)-Michelson interferometer and also greatly simplifies the optical alignment of the system.

### 5. Conclusions

Because the arrival on the earth of the Gravitational Wave (GW) typically results in a very small dilation/compression change of any physical lengths of all existing objects, GW detection necessarily requires that a very, very large "effective length" $L_{eff}$ must be measured, via the only existing GW-invariant "meter", the light wavelength for free-space propagation, by any very, very localized detection apparatus. This is indeed the oxymoron lying at the root of all problems affecting this peculiar kind of quantum measurement. In order to solve these problems in US, France and Italy in 1980 imaginative, bold people started the "adventure" of optical Michelson Interferometry (MIF). As well known, this adventure, after about fifty years of intense struggle recently attained a well-deserved success (1). In 1980 no other sensible solutions were in sight, apart from previous John Weber massive-antenna rather inconclusive solutions, that about at that time were dismissed (10). Optical Fiber (OF) technology was in its infancy in 1980. However, only a few decades later with the advent of the world wide web and the starting of the oceanic cable long-range communications the situation started to change dramatically. Today this technology has reached a high degree of sophistication by offering a rich scenario of technical and yet unexplored scientific solutions. The proposal presented in this work is aimed to take advantage of this progress by offering a solution to GW detection that is far simpler and conceivably far less expensive that the (GW-MIF). As shown in the present text, a large part of the heavy noise contributions that plague the (MIF) operation together with the contemporary need for a very accurate (FP) mirror alignment in a noisy environment imposed by the attainment of a very large (IF) "Finesse" (an incredible challenge indeed!), are absent in our system. There is no need for soil excavations or building constructions. Furthermore, and most important in the present context, this proposal implies the possibility of a complete survey of the skies by an easy orientation of the Antennas. We believe that the present solution can open a new set of scientific opportunities in the novel field of GW Astronomy.

**Figure Captions**

1) GW effect of the two polarizations $h_+$ and $h_x$ over a circular distribution of masses.
2) Light transmission loss in a single – mode OF. In this work the most convenient Telecom C-Band is selected.
3) Layout of the Proposed Fiber Optics Interferometer in the X-Y plane. The two equal Antennas are clamped by motorized mechanical devices to a single external tower allowing independent orientations towards the skies in any possible direction.
4) Layout of the Proposed Fiber Optics Antenna in the X-Z plane. The metal slab is tightly enfolded by OF windings, which are parallel to the X axis.
5) Layout of the Proposed Fiber Optics Antenna in the X-Y plane.
6) Transverse section of two typical Hollow-Core Fibers. The typical structure of the HC-NANF is shown in the right side. NO-BENDING fiber Configuration.
7) Layout of the typical configuration for long – range OF communication
8) Structure of a typical EDFA.
9) Quantum levels of the Yitterbium-Erbium system for high power amplification (YEDFA).
10) Layout of our system in a conventional MI–IF configuration. The end – mirrors for the Antennas can be either external or "internal" i.e. belonging to the "core" of the OF structure itself. Optionally external or "internal" Fabry – Perot mirrors can be added at the input of the Antennas.

# 1) GW effect over a circular distribution of masses.

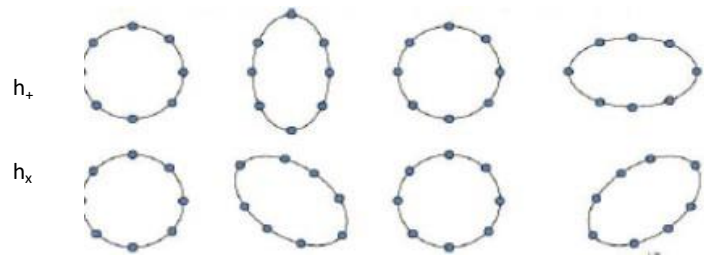

$h_+$

$h_\times$

---

# 2) Transmission loss in a single-mode fiber

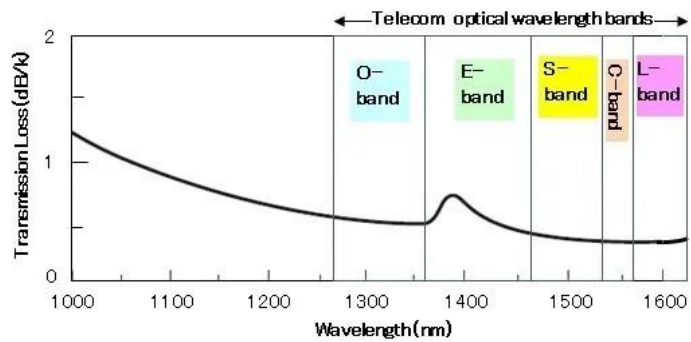

---

## 3) OF Mach-Zehnder Interferometer

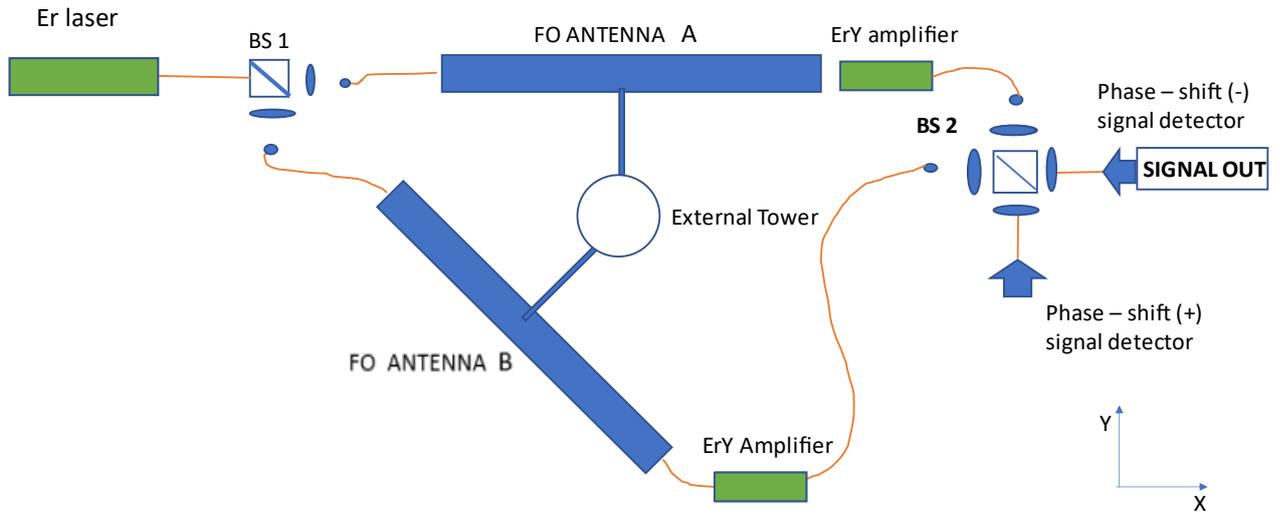

## 4) Fiber Optics Antenna (x-z)

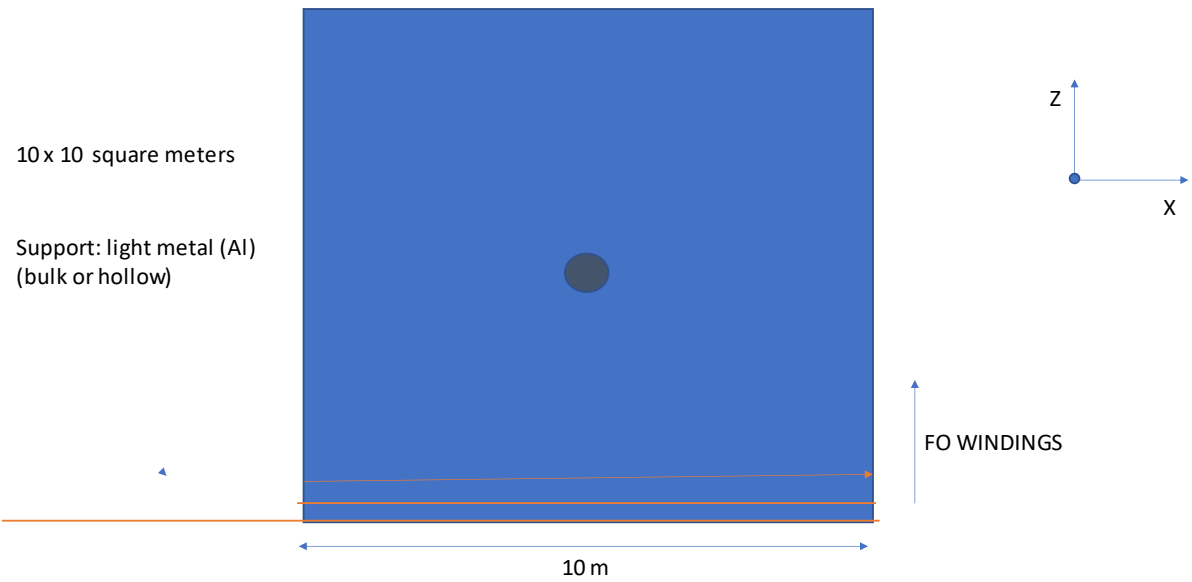

## 5) FiberOpticsAntenna (x – y)

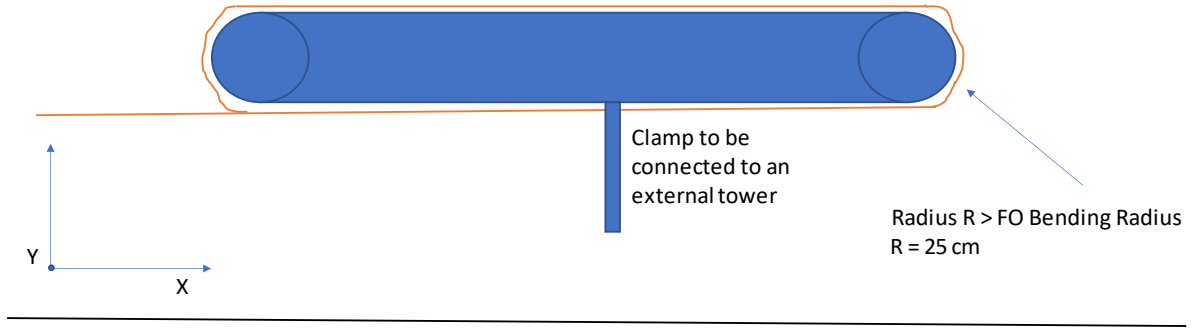

Clamp to be connected to an external tower

Radius R > FO Bending Radius
R = 25 cm

Y
X

## 6) Transverse section of HC Fibers

HC-NANF

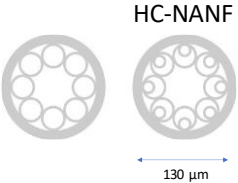

130 µm

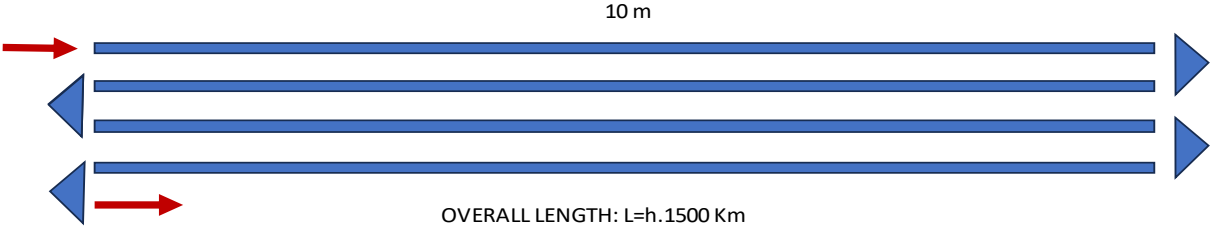

10 m

OVERALL LENGTH: L=h.1500 Km

**NO FIBER BENDING**

# 7) OF long range communication

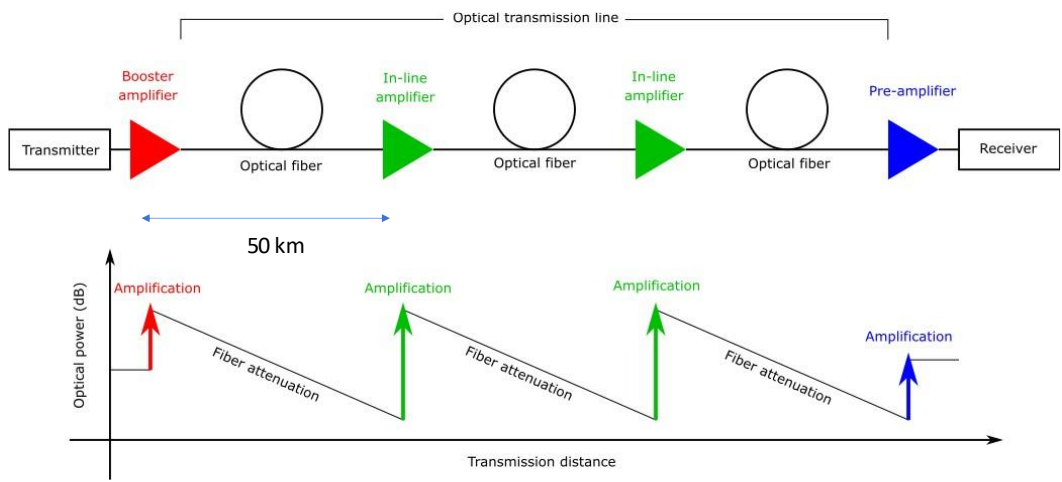

# 8) Erbium Doped Fiber Amplifier

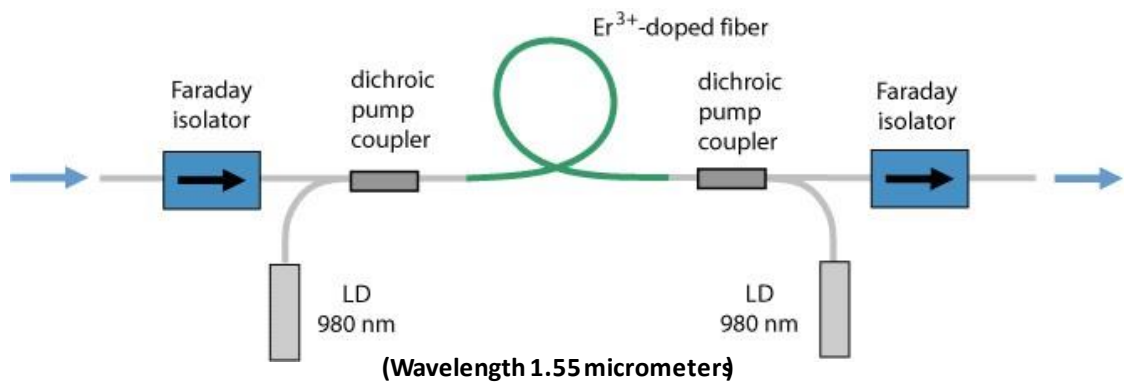

**(Wavelength 1.55 micrometers)**

## 9) Levels of the Ytterbium-Erbium system

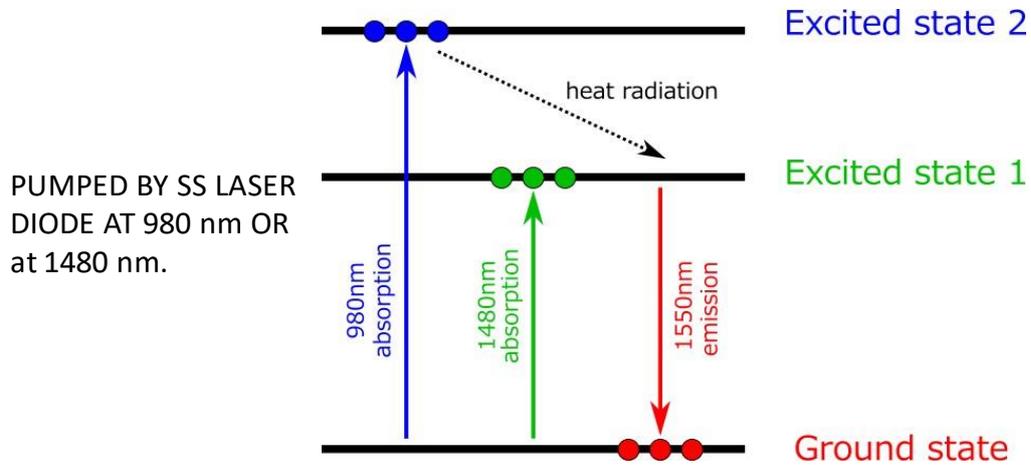

PUMPED BY SS LASER DIODE AT 980 nm OR at 1480 nm.

## 10) Michelson OF Interferometer

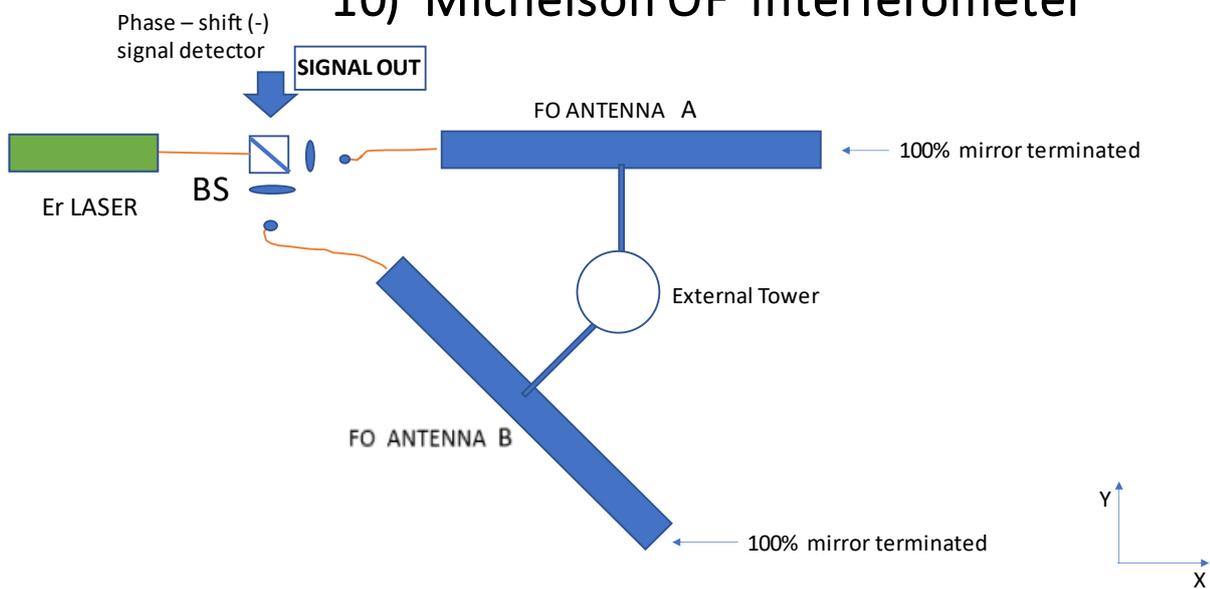